\documentstyle[prd,aps]{revtex}
\begin{document}
\draft
\input epsf

\def\simlt{\stackrel{<}{{}_\sim}}
\def\simgt{\stackrel{>}{{}_\sim}}

\twocolumn[\hsize\textwidth\columnwidth\hsize\csname
@twocolumnfalse\endcsname

\title{Standard Model Neutrinos as Warm Dark Matter}
\author{Gian F.\ Giudice$^{(1)}$, Edward W.\ Kolb$^{(2)}$, 
Antonio Riotto$^{(3)}$, Dmitry V.\ Semikoz$^{(4,5)}$ 
and Igor I.\ Tkachev$^{(1,5)}$}

\address{$^{(1)}${\it CERN Theory Division, CH-1211 Geneva 23, Switzerland}}
\address{$^{(2)}${\it NASA/Fermilab Astrophysics Center, Fermi
     National Accelerator Laboratory, Batavia, Illinois \  60510-0500,\\
     and Department of Astronomy and Astrophysics, Enrico Fermi Institute, \\
     The University of Chicago, Chicago, Illinois \ 60637-1433}}
\address{$^{(3)}${\it Scuola Normale Superiore, Piazza dei Cavalieri 7, 
I-56126 Pisa, Italy}}
\address{$^{(4)}${\it Max Planck Insitut f\"ur Physik, F\"ohringer Ring 6,
80805 M\"unchen, Germany}}
\address{$^{(5)}${\it Institute for Nuclear Research of the 
Academy of Sciences of Russia, Moscow 117312, Russia }}

\date{December, 2000}
\maketitle
\begin{abstract}
Standard Model neutrinos are not usually considered plausible dark
matter candidates because the usual treatment of their decoupling
in the early universe implies that their mass must be sufficiently
small to make them ``hot'' dark matter.  In this paper we show that
decoupling of Standard Model neutrinos in low reheat models may result in
neutrino densities very much less than usually assumed, and thus their
mass may be in the keV range. Standard Model neutrinos may therefore be
warm dark matter candidates.
\end{abstract}
\vskip2pc]

\section{Introduction} 
Cosmological dark matter in the form of neutrinos with masses in the eV range
is the quintessential example of hot dark matter (HDM)
\cite{review}. Indeed, for a standard relic that decouples from the 
surrounding thermal bath when still relativistic, the current abundance 
can be easily estimated to be \cite{book}
\begin{equation}
\label{st}
\Omega_X h^2\simeq 78\; \left(\frac{g_X}{g_*(T_D)}\right)
\left(\frac{m_X}{{\rm keV}}\right),
\end{equation}
where $g_X$ is the number of degrees of freedom of the particle and $g_*(T_D)$
is the total effective number of relativistic degrees of freedom at the
decoupling temperature $T_D$. Relativistic standard model neutrinos decouple
from chemical equilibrium at a temperature of a few MeV, when $g_*(T_D)\simeq
10.75$.  Since current observations indicate that the dark matter density
amounts to approximately $\Omega_X\simeq 0.3$, one usually concludes that the
mass of standard model neutrinos cannot be larger than about $30h^2$\,eV
\cite{book,gersht}.
                                                                  
Since neutrinos must be light in order to avoid overclosing the universe, they
were moving at nearly the speed of light at redshift $z\sim10^6$ when the
cosmic horizon first encompassed $10^{12}M_\odot$, the amount of dark matter
contained in the halo of a large galaxy like the Milky Way.  This implies that
the free streaming of light neutrinos destroyed any fluctuations smaller than
that of a supercluster (about $10^{15}M_\odot$). Cosmological structure forms
in a light-neutrino dominated universe in a top-down scenario, in which
superclusters of galaxies form first, with galaxies and clusters forming
through a process of fragmentation.  However, in this scenario galaxies form
too late and their distribution is much more inhomogeneous than observations
indicate.  An even more serious problem is that when normalized to the low
amplitude of the cosmic microwave background fluctuations detected by the COBE
satellite \cite{Smoot92}, the HDM spectrum is only beginning to reach
nonlinearity at the present epoch and the free-streaming wavelength cutoff in
the amplitude of the spectrum should be considerably smaller to form any
structure by the present.  Because of these difficulties, the idea of standard
model neutrinos as dark matter has been gradually abandoned in favor of
scenarios where most of the mass in the universe is in the form of cold dark
matter (CDM).

The purpose of this paper is to show that in low-reheat cosmologies, standard
model neutrinos may play the role of warm dark matter (WDM), avoiding the
pitfalls of HDM, and may be responsible for the cosmological structures we
observe.

The key point of our observation is that the limit to the contribution of
neutrinos to the present density need not constrain neutrino masses to be in
the eV region. On the contrary, neutrino number densities in low-reheat models
may be more than an order of magnitude lower than usually assumed, and their
mass, $m_\nu$, more than an order of magnitude higher, perhaps in the keV
range. Fluctuations corresponding to sufficiently large galaxy halos with
masses around $10^{11}M_\odot$ may survive free streaming and standard model
neutrinos may act as WDM. Before launching into more details, let us see why
standard model neutrinos may be heavier than usually thought.

\section{On the neutrino density}
The traditional computation of the abundance of standard model neutrinos in the
early universe is based on the simple, but {\it untested,} assumption that
light neutrinos were in chemical equilibrium at temperatures larger than
$T_D$. In other words, in the standard cosmological bound, $m_\nu \simlt
30h^2$\,eV, it is tacitly assumed that the universe had gone through a
radiation-dominated phase with temperatures larger than about an MeV with
active neutrinos in equilibrium.  The assumption of an initial condition of
neutrinos in thermal and chemical equilibrium in a radiation-dominated universe
is then equivalent to the hypothesis that the maximum temperature obtained
during the (last) radiation-dominated era, which will refer to as the reheating
temperature $T_{RH}$, is much larger than the decoupling temperature.  The fact
that we have no physical evidence of the radiation-dominated era well before
the epoch of nucleosynthesis is a simple, but crucial, point.  It was shown in
Ref.\ \cite{kks} that the reheating temperature of order 1 MeV is still
compatible with light-element nucleosynthesis. A priori, one should consider
$T_{RH}$ as an unknown quantity that can take any value as low as 1 MeV.

It is usually assumed that the radiation-dominated era commences after a period
of inflation, and that the cold universe at the end of inflation becomes the
hot universe of the radiation-dominated era in a process known as reheating.
The reheating process need not be instantaneous. On the contrary, before the
radiation-dominated phase there may have been a prolonged phase during which
the energy density of the universe was dominated by some component other than
radiation. This component is often represented by a coherent oscillating field
such as the inflaton field, but one could just as easily imagine that the
universe is dominated by some unstable massive particle species. During
reheating there is a slow formation of a thermal bath of relativistic
particles. The temperature of this thermal bath has a peculiar behavior
\cite{book}. It reaches a maximum temperature $T_{\rm max}\sim T_{RH}\left(
H_I^2 M_{Pl}^2/T^4_{RH}\right)^{1/4}$ ($H_I$ is the value of the Hubble rate at
the beginning of the reheating process) and then has a less steep dependence on
the scale factor $a$ than in the radiation-dominated era, $T\sim
a^{-3/8}$. During this phase entropy is continuously created (the universe is
reheating!)  and the Hubble rate scales like $H\sim
[g_*(T)/g_*^{1/2}(T_{RH})](T^4/ T_{RH}^2 M_{Pl})$.

At a given temperature, the expansion is faster for smaller reheat
temperatures. When the temperature decreases to $T_{RH}$, the universe
enters the radiation phase, and one recovers the more familiar Hubble
law, $H\sim T^2/M_{Pl}$.

Let us now assume that the largest temperature of the universe during the
radiation-dominated phase is very small, of the order of a few MeV.  Since
neutrinos have only weak interactions, it is very difficult for the thermal
scatterings during the reheating stage to generate standard model neutrinos
through processes like $e^+ e^-\rightarrow \nu\bar\nu$ and to bring neutrinos
into chemical equilibrium. Furthermore, decreasing the reheat temperature
increases the rate of the expansion of the universe, making it more and more
difficult for the weak interactions to bring the neutrinos to chemical
equilibrium. Therefore, if the reheating temperature is small enough, standard
model neutrinos produced during the reheating stage {\it never} go into
chemical equilibrium. In this case, neutrinos are present in the thermal bath
at the beginning of the radiation-dominated phase, but they have a number
density $n_\nu$ that is {\it smaller} than the equilibrium number density
\cite{gkr}. This simple argument shows that the present abundance of neutrinos
may be much smaller than predicted assuming that the largest temperature of the
radiation-dominated universe was much larger than a few MeV.  Thus, standard
model neutrinos heavier than about $30 h^2$\,eV are perfectly compatible with
cosmology.

The above expectation was confirmed in Ref.\ \cite{gkr}, where the effective
number density of neutrinos was computed by solving the corresponding Boltzmann
equation obtained under the assumptions of Maxwell--Boltzmann statistics and
local thermodynamic equilibrium in the calculation of the thermal averaged
cross section.  This amounts to assuming that electrons have an equilibrium
Boltzmann distribution function and to neglect all the Pauli blocking factors.
In this paper we have performed a more refined and correct computation by
solving directly the kinetic equations for the neutrino phase-space
distribution $f_{\nu}(p,t)$.

We will use the numerical code developed in Ref.\ \cite{dhs97}, where
the collision integrals are analytically reduced to two dimensions
\cite{ST}.  The kinetic equations for neutrinos have the form
\begin{equation}
        \frac{\partial f_{i}(p_1,t)}{\partial t}
        - H(t)p_1\frac{\partial f_{i}(p_1,t)}{\partial p_1}=I_{i,{\rm coll}}, 
        \label{kin1}
\end{equation}
where the collision integral $I_{\rm coll}$ is dominated by two-body reactions
$1+2 \rightarrow 3+4$,
and is given by the expression
\begin{eqnarray}
        I_{\rm coll} & & 
= {S\over 2E_1}\sum \int {d^3 p_2 \over 2E_2 (2\pi)^3}
        {d^3 p_3 \over 2E_3 (2\pi)^3}{d^3 p_4 \over 2E_4 (2\pi)^3}(2\pi)^4
        \nonumber \\ && 
        \delta^{(4)} (p_1+p_2-p_3-p_4) F(f_1,f_2,f_3,f_4)
        |{\cal M}|^2_{12\rightarrow 34}.
        \label{icoll}
\end{eqnarray}
Here $F = f_3 f_4 (1-f_1)(1-f_2)-f_1 f_2 (1-f_3)(1-f_4)$, $|{\cal
M}|^2$ is the square of the weak-interaction amplitude summed over
spins of all particles except the first one, and $S$ is the
symmetrization factor which includes $(1/2!)$ for each pair of
identical particles in the initial and final states and a factor of 2
if there are two identical particles in the initial state. Finally,
there is a summation over all possible sets of leptons 2, 3, and 4.
Notice that a similar approach was taken in Ref.\ \cite{kks}, even though
there the interactions among neutrinos as well as the electron mass
were neglected and electrons were assumed to have a Boltzmann
distribution. (These approximations allow the reduction of the
collision integrals in the kinetic equations to one-dimensional
integrals, thus simplifying the numerical calculations.)  Furthermore,
in Ref.\ \cite{kks} the emphasis was on the impact of very low
reheating temperatures on standard big-bang nucleosynthesis, and
therefore neutrinos were taken to be massless.

We have assumed that reheating is due to the decay into light states
of a particle $\phi$, which might be the inflaton field, a modulus
field, or any unstable particle which dominated the energy density of
the universe before the radiation-dominated phase.  The time evolution
of the $\rho_{\phi}$ energy density is given by
\begin{equation}
        \frac{d\rho_{\phi}}{dt}  =  -\Gamma\rho_{\phi} -3H\rho_{\phi},
        \label{eq:rho_phi}
\end{equation}
where $\Gamma$ is the decay rate of the $\phi$ field, which may be expressed in
terms of the reheat temperature $T_{RH}$ as $\Gamma = 3H =
3(T_{RH}^2/M_{Pl})[8\pi^3g_*(T_{RH})/90]^{1/2}$ where $g_*(T_{RH})$ is the
number of relativistic degrees of freedom at $T_{RH}$.  Notice that in Refs.\
\cite{kks,gkr}, the value of $g_*$ in the $(\Gamma$--$T_{RH})$ 
relation was fixed to 10.75.  This is not necessarily the actual value of
$g_*(T_{RH})$; we will return to this point below.  Also, in Ref.\ \cite{gkr},
the definition of $T_{RH}$ is $\Gamma=H$, rather than $\Gamma=3H$.

We imposed the covariant energy conservation $\dot{\rho}(t) = -3 H (\rho + P)$,
where $\rho$ is the total energy density is
\begin{eqnarray}
        \rho & = & \rho_{\phi}(t) + {\pi^2 T^4_\gamma\over 15}  + 
        \frac{2}{\pi^2} \int dq \,q^2
        \frac{\sqrt{q^2 + m^2_e}}{\exp {(E/T_\gamma)} +1} \nonumber\\
        & & + \frac{1}{\pi^2} \int dq \, q^3 f_{\nu_e}(q) + \frac{2}{\pi^2} 
        \int dq \,q^3 f_{\nu_\mu}(q) ,
\label{rho}
\end{eqnarray}
and a similar expression holds for the pressure $P$.  The initial density of
the scalar field $\rho_\phi$ is not relevant since it just defines the initial
time of the evolution.

\begin{figure}
\centering\leavevmode\epsfxsize=3.4in\epsfbox{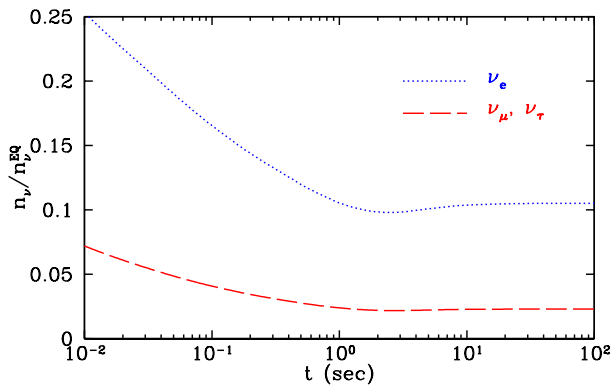}
\caption{The relative neutrino densities as functions of time for 
$T_{RH}= 1$ MeV.} \label{fig:n_1MeV.ps}
\end{figure}

At the beginning of the evolution, the electromagnetic interactions are much
faster than the neutrino interaction rates and therefore, before 
neutrinos are produced,
the plasma reaches temperatures higher than the final reheating temperature
$T_{RH}$ \cite{book,gkr}. Later, neutrinos $\nu_i$ of a given flavor start
being produced by electron-positron annihilations ($e^+e^-\rightarrow
\nu_i\bar{\nu}_i$) and by neutrino-(anti)neutrino annihilations
($\nu_j\bar{\nu}_j\rightarrow \nu_i\bar{\nu}_i$ with $i\neq j$). While tau- and
muon-neutrinos are produced only by neutral current interactions,
electron-neutrino production has a contribution from charged current
interactions as well. However, since the present bound on $m_{\nu_e}$ is in the
eV range, we will not be interested in $\nu_e$'s as dark matter. At late times,
the neutrino distribution reaches some dynamical shape which differs from the
equilibrium one. We have checked that the neutrino abundances follow the same
evolution curve regardless of their initial abundances and the final neutrino
distribution is insensitive to the initial conditions.  The time dependence of
neutrino density (at late times) is shown in Fig.\ \ref{fig:n_1MeV.ps}.

\begin{figure}
\centering\leavevmode\epsfxsize=3.4in\epsfbox{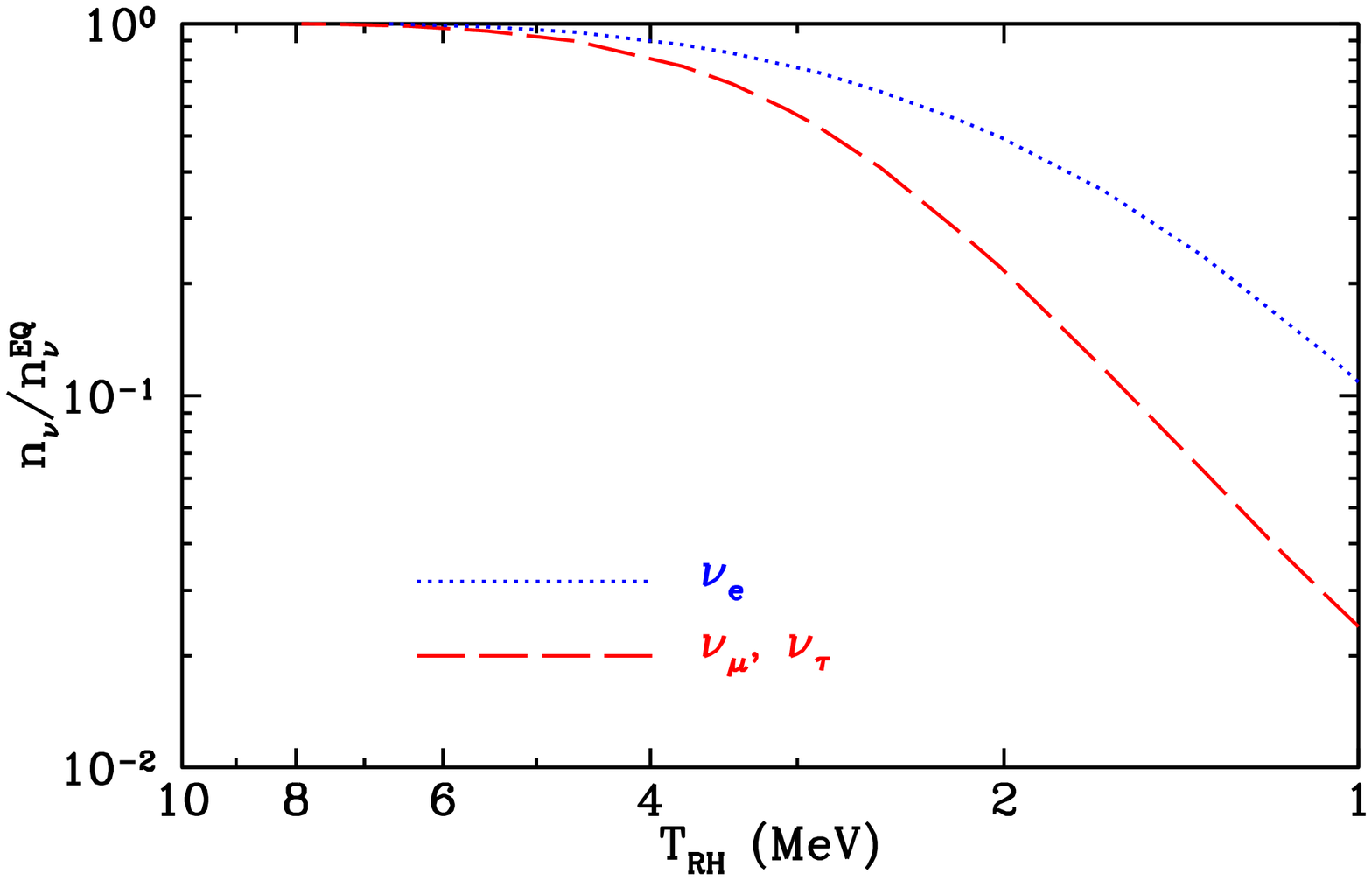}
\caption{The relative number density of different neutrino species 
to the equilibrium neutrino number density as a function of the 
reheating temperature.} \label{fig:n_T}
\centering\leavevmode\epsfxsize=3.4in\epsfbox{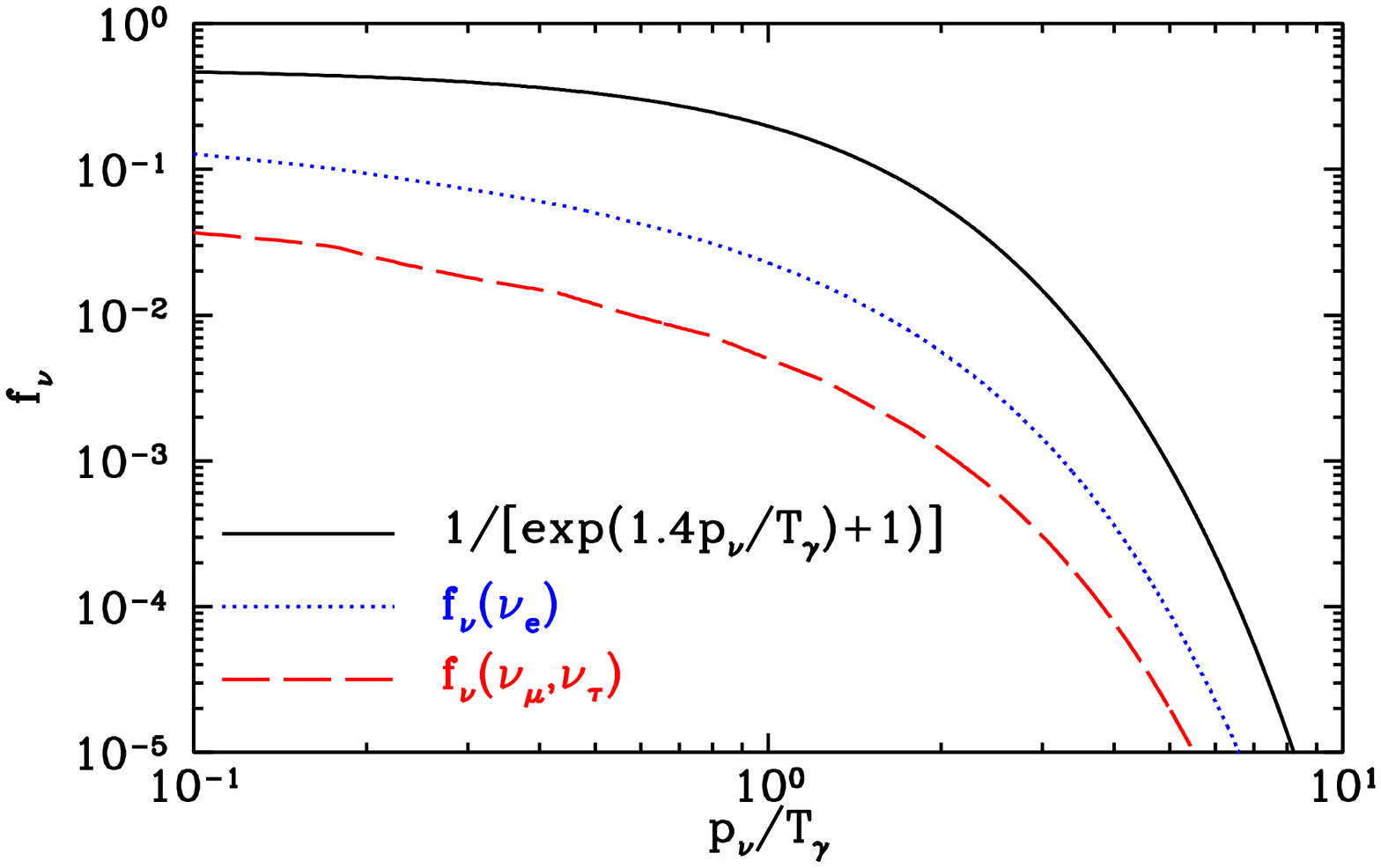}
\caption{The final distribution functions for $\nu_e$, $\nu_\mu$ 
and $\nu_\tau$ for $T_{RH}= 1$ MeV as function of the dimensionless 
momentum $p_\nu/T_{\gamma}$.  The dotted line corresponds to the 
standard model neutrino distribution function.}
\label{fig:distr}
\end{figure}

The final tau- and muon-neutrino number density (they are the same) normalized
to the neutrino number density in equilibrium $n_\nu/n_\nu^{{\rm EQ}}$ (where
$n_\nu^{{\rm EQ}} = 2\xi(3)T_\nu^3/\pi^2$) is plotted in Fig.\ \ref{fig:n_T} as
a function of $T_{RH}$. This ratio becomes smaller than unity for $T_{RH}$
smaller than about 8 MeV, signalling a departure from equilibrium. The final
momentum distribution function for $T_{RH}=1$ MeV is plotted in Fig.\
\ref{fig:distr}. For comparison, we have plotted the equilibrium
neutrino distribution function taking into account the fact that the
neutrino temperature is a factor $1.4$ times smaller then the photon
temperature $T_\gamma$ due to $e^{\pm}$ annihilation after neutrino
decoupling.

What is relevant is that the abundance of tau- and muon-neutrinos is
about a factor of $2.7\times 10^{-2}$ smaller than the standard
abundance for reheating temperatures around 1 MeV.  
This in turn implies that the
standard upper bound of $30 h^2$\,eV on $m_\nu$ no longer applies and
the abundance of tau- and muon-neutrinos for $1\ {\rm MeV} \simlt
T_{RH} \simlt 3$ MeV can be expressed as
\begin{equation}
        \Omega_{\nu_{\tau}}h^2=
        \Omega_{\nu_{\mu}}h^2=\left(\frac{m_\nu}{4\ {\rm keV}}\right)
        \left(\frac{T_{RH}}{1\ {\rm MeV}}\right)^{3},  \label{onu}
\end{equation}
where we used a power-law approximation $n_\nu/n^{{\rm EQ}}_\nu 
\approx 0.024 (T_{RH}/1\ {\rm MeV})^3$. For larger reheat
temperatures, $3\ {\rm MeV}\simlt T_{RH}\simlt 8\ {\rm MeV}$, one needs a
slightly more complicated expression to fit $n_\nu$: $n_\nu/n^{{\rm EQ}}_\nu
\simeq 0.44 \; {\rm tan}^{-1}
\left[(T_{RH}({\rm MeV})-2.6)/1.17\right] + 0.43$.

\begin{figure}
\centering\leavevmode\epsfxsize=3.4in\epsfbox{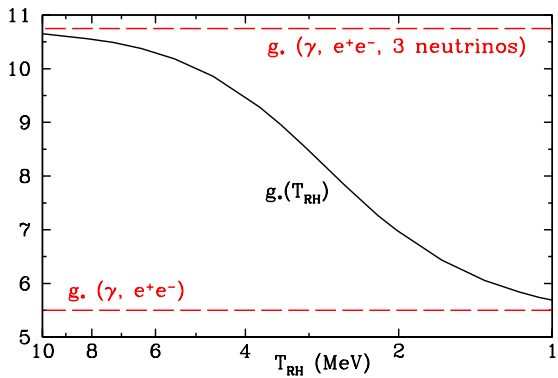}
\caption{The actual value of $g_*$ at $T_{RH}$ as a function of $T_{RH}$.}
\label{fig:gstar}
\end{figure}

As mentioned above, the value of $g_*$ in the $(\Gamma$--$T_{RH})$ relation is
actually a function of $T_{RH}$.  The value of $g_*(T_{RH})$ provides an
indication of the relative abundance of neutrinos in the thermal bath.  The
value of $g_*(T_{RH})$ as a function of $T_{RH}$ is given in Fig.\
\ref{fig:gstar} and allows to relate the
decay rate to the  reheat temperature, 
$\Gamma^{-1} =({\rm MeV}/T_{RH})^2 g_*(T_{RH})^{-1/2}1.6$~sec
$\simeq 0.56 ({\rm MeV}/T_{RH})^{2.16}$ sec.  

Our result in Eq.\ (\ref{onu}) is slightly smaller than the relic neutrino
abundance computed in Ref.\ \cite{kks}. For instance, Fig.\ 2 of Ref.\
\cite{kks} shows a value of $\Omega_\nu h^2$ a factor of 1.6 larger than what
is given by Eq.\ (\ref{onu}), for $T_{RH}=2$ MeV.  
The calculation in Ref.\ \cite{kks} makes the approximation of {\it (i)}
neglecting the electron mass in the collision integrals, {\it (ii)} assuming
a Boltzmann distribution for the electrons, {\it (iii)} neglecting
$\nu$--$\nu$ interactions. We find that the first approximation is well
justified, but the other two give errors of up to  30\% (depending on the 
value of $T_{RH}$), which however are opposite in sign and roughly cancel
each other.
The numerical discrepancy is largely
explained by the fact that the authors of Ref.\ \cite{kks} fix
$g_*(T_{RH})=10.75$, and therefore their result should be multiplied by a
correction factor $[10.75/g_*(T_{RH})]^{3/4}$. For the same reason, the limit
on $T_{RH}$ derived in Ref.\ \cite{kks} from nucleosynthesis should also be
rescaled by a factor $[10.75/g_*(T_{RH})]^{1/4}$. With this procedure we find
that nucleosynthesis gives slightly more stringent lower bounds on $T_{RH}$:
$T_{RH}>1.2$ MeV (at 68\% CL) and $T_{RH}>0.8$ MeV (at 95\% CL).

The result of Eq.\ (\ref{onu}) is similar to that found in Ref.\ 
\cite{gkr} (allowing
for the differing definition of $T_{RH}$ mentioned above).  This small
difference traces to the assumptions made in Ref.\ 
\cite{gkr} of Maxwell--Boltzmann
statistics and local thermodynamic equilibrium in the calculation of the
thermal averaged cross section.

\section{Standard model neutrinos as warm dark matter} 
CDM reproduces the observable universe at large scales, but appears to be in
conflict with observations on sub-galactic scales. CDM produces too many dwarf
galaxies and overdense galactic cores compared to observations.  The resolution
of this difficulty has been searched for along different routes.  Many
suggestions can be united under recipe of reducing the power on small scales.
In the WDM scenario, this reduction occur naturally via the mechanism of free
streaming. The comoving smoothing scale can be estimated as the comoving
horizon at matter-radiation equality times the {\it rms} velocity of dark
matter particles at that time \cite{fsl} 
\begin{equation}
R \approx 0.2 \left({\Omega_\nu h^2}\right)^{1/3} 
\left({\rm keV}/m_X\right)^{4/3} {\rm Mpc}.  \label{sl}
\end{equation}

Interestingly, $\Omega_X$ also takes correct value within the same range of
$m_X$ if the reheating temperature is of order 1 MeV [see Eq.\ (\ref{onu})].
This numerical coincidence deserves further study and detailed numerical
investigation. 
\begin{figure}
\centering\leavevmode\epsfxsize=3.4in\epsfbox{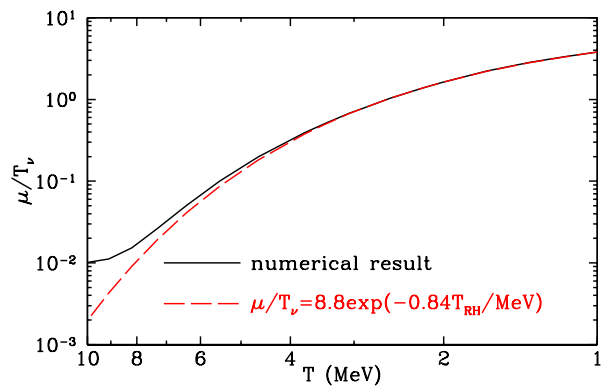}
\caption{The effective chemical potential as a function of 
reheat temperature.}
\label{fig:mu}
\end{figure}

Encouraging preliminary results were obtained in Ref.\ \cite{BOT}. To aid such
studies, we parameterize our final distribution function of neutrinos as a
thermal distribution with some effective ``temperature'' and a ``chemical
potential.''  The difference between the effective temperature and the
standard-model neutrino temperature is insignificant (only a few percent) and
the modification to the distribution function can be attributed to the chemical
potential, which is shown in Fig.\ \ref{fig:mu}.

At large $T_{RH}$ the chemical potential approaches the standard-model value
$\mu=0.01 T_\nu$. At small $T_{RH}$ it can be fitted as $\mu/T_\nu=8.8 \,
\exp(-0.84 T_{RH}/$MeV).

To conclude, the combination of a standard model neutrino of mass of a few keV
and a reheat temperature of about an MeV will result in neutrinos as candidate
WDM.  Assuming $\Omega_\nu=0.3$ in one species of neutrino and $h=0.65$, then
$R=0.4 (T_{RH}/{\rm MeV})^4$\ Mpc.  As discussed above, any reheat temperature
above about an MeV is consistent with BBN. 

We want to stress that the hypothesis of neutrinos as WDM is not inconsistent
at present with the recent data on atmospheric and solar neutrino
anomalies. Atmospheric neutrino data could be accomodated by oscillations
between quasi-degenerate $\nu_\mu$ and $\nu_\tau$ states and in such a case
solar neutrino observations would require $\nu_e$ to oscillate into a sterile
state. Alternatively, the solar neutrino data might be explained by conversion
of $\nu_e$ into an active state and the atmospheric neutrino deficit by the
conversion of $\nu_\mu$ into a sterile state.

\acknowledgements{The work of E.W.K.\ was supported in part by NASA
(NAG5-7092).  }

\end{document}